\documentclass[12pt,a4paper]{article}
\usepackage[latin1]{inputenc}
\usepackage[T1]{fontenc}
\usepackage[english]{babel}
\usepackage{a4,array,latexsym}

\newcommand{\bc}{\begin{center}}
\newcommand{\ec}{\end{center}}
\newcommand{\bd}{\begin{displaymath}}
\newcommand{\ed}{\end{displaymath}}
\newcommand{\be}{\begin{equation}}
\newcommand{\ee}{\end{equation}}
\newcommand{\ba}{\begin{array}}
\newcommand{\ea}{\end{array}}
\newcommand{\bea}{\begin{eqnarray}}
\newcommand{\eea}{\end{eqnarray}}
\newcommand{\bt}{\begin{tabular}}
\newcommand{\et}{\end{tabular}}

\newcommand{\bp}{\begin{picture}}
\newcommand{\ep}{\end{picture}}
\newcommand{\bfi}{\begin{figure}}
\newcommand{\efi}{\end{figure}}
\begin{document}


\title{Discussion on the Cosmological Vacuum Energy}

\author{H.B.~Nielsen\footnote{hbech@nbi.dk} 
\\
\itshape{The Niels Bohr Institute, Copenhagen, Denmark}
\\[3mm]
C.~Bal\'azs\footnote{csaba.balazs@monash.edu} 
\\ 
\itshape{School of Physics, Monash University, Melbourne, Australia} 
}

\date{2011 Aug 14}

\maketitle

\begin{abstract}
The present discussion contribution is some remarks concerning and review 
of the proposal by one of us to explain the cosmological constant by 
a/the principle of entropy. Used without further comment this 
principle of entropy could easily lead  to untrustable 
{\em nonlocalities}, but
taking into account that the long range correlations are rather to 
be understood as due to initial condition set up the model for 
the cosmological constant being small by one of us becomes 
quite viable.
\end{abstract}

\newpage

\thispagestyle{empty}

\section{Introduction}

In a recent paper one of us \cite{Balazs:2006sd} proposed to explain the 
smallness of the cosmological vacuum energy based on the energy limit that 
general relativity imposes on any given volume.  According to Einstein's theory 
the maximal amount of energy in a 3-dimensional volume scales with the linear 
size rather then the volume itself.  This energy limit can be equivalently 
formulated as an upper limit on the entropy contained in the volume, which 
latter is known as the holographic entropy bound.\footnote{See Ref. 
\cite{Balazs:2006sd} for detailed literature on the subject.}  The entropy bound 
is well known, for example, for Schwarzschild black holes: the entropy contained 
within the horizon cannot exceed the quarter of the surface area (in Planck 
units).

>From this entropy, or equivalently energy, bound it follows that there must 
exist an ultraviolet cut-off for fields in the region inside any volume 
\cite{Balazs:2006sd}.  Simply put: the entropy (energy) in the given volume 
cannot exceed the maximal entropy (energy) of a black hole that fills the given 
volume.  With such a cut-off the zero point energy of any fields, that is the 
vacuum energy, inside any volume becomes restricted.  This in turn limits the 
cosmological vacuum energy which otherwise would contribute to dark energy (or 
the cosmological constant) an enormous amount.  Such a restriction is not only 
suitable to reduce dark energy (cosmological constant), but as explained in Ref. 
\cite{Balazs:2006sd} it will ensure that the theoretical prediction of the 
cosmological vacuum energy will exactly match that of the experimentally 
measured value.

However, it is non-trivial to understand how such an energy cut-off is 
implemented in quantum field theory.  The most naive implementation of such a 
cut-off would be to restrict the individual degrees of freedom within a given 
volume independently not to exceed their average energy.  Unfortunately, this 
cut-off would lead to an energy limit that scales with the volume rather than 
the linear size.  Moreover, in case of the cosmological vacuum energy this is 
experimentally excluded since locally we observe higher energy density systems 
in our Universe than its average dark energy density.  Thus the cut-off has to 
vary and has to be non-trivially correlated between degrees of freedom of the 
Universe.  This would allow for the existence of small regions with high energy 
densities while the rest could compensate such that the average never exceeds 
the ratio of the maximal allowed energy and the volume.

But this sort of cut-off raises another question: How can such a correlated cut-
off be consistent with the locality of quantum field theory?  If no signal 
propagates faster than the speed of light, can potentially distant parts of the 
volume compensate for each other?  Naturally, whether such a cut-off is 
consistent with locality depends on the details of the implementation of the 
cut-off itself.  Most importantly, how the long range correlation between the 
allowed energy in one region depends on what goes on or what is allowed in an 
other region.  But if the cut-off at one region depends on what goes on at an 
other, potentially distant, region {\em at the same time}, then locality can 
definitely get into trouble!  So it can at best be a tolerable correlation of 
the cut-off at one locality with what went on somewhat earlier around the 
Universe in remote regions, otherwise causality is threatened.

To implement such a cut-off while saving locality, one may think in two different ways:

a) One could accept that locality is not necessarily a good principle and the 
solution necessitates ``new physics''.  But then one is up to theories like the 
``complex action theory'' proposed by Ninomiya and the other one of the present 
authors(HBN)\cite{complex}.

b) An alternative solution is the use of some cosmic censorship assumption 
such as the non-existence of ``white holes'', that is time-reversed 
black holes.  Such an assumption is needed anyway, to maintain the entropy bound \cite{bousso}. 

It appears that in quantum field theory the entropy bound holds only if either 

\begin{itemize}

\item the cut-off is strangely correlated between the degrees of freedom, as suggested by \cite{Balazs:2006sd}, or 

\item the limitation of the number of states is not just a limitation
due to the cut-off of the theory but due e.g. to some special initial
condition.  And as an example of the latter - one of us would say 
more reasonable type of state limitation for the application in question - 
the cosmic censorship comes in.

\end{itemize}

In Section \ref{trouble} we discuss the problems related to the argument that the cosmological vacuum energy is limited by the entropy bound.  In Section \ref{resolution} we put forward an idea of how a cut-off based on the entropy bound could be interpreted or replaced by a cosmic censorship based philosophy.  This latter could, at least in a certain sense, be free of the problems with locality or causality.  Finally we conclude and look out in the conclusion section \ref{conclusion}.

\section{Trouble for the Entropy Principle}
\label{trouble}

In a physical system obeying the laws of thermodynamics the extensive 
thermodynamical variables, such as energy and entropy, typically scale with the 
volume containing the system.  Since in quantum systems the energy, in turn, 
typically scales with the number of degrees of freedom, the latter is usually 
thought to grow with the volume.  Considering a system of fields defined within 
a volume, without any special restrictions on the degrees of freedom, it is 
clear that the number of states can grow as an exponential of the volume.  
In field theory in any local region of space the energy density can reach that 
of the highest energy accelerators and beyond, and the number of degrees of 
freedom in a given volume is unlimited.  Because of this field theory does not 
respect the Bekenstein-Hawking entropy bound or the Schwarzschild energy limit. 
As we saw before, the entropy bound as a naive, uncorrelated cut-off on any 
given volume is out of question in field theory.  The entropy and energy bounds 
can only be consistent with field theory, they can only allow reaching energy 
densities well tested in science and in daily life, if the corresponding cut-off 
is highly correlated between the degrees of freedom.  Thus the nature of the cut-off is such that it imposes a strong restriction on the allowed states.


About a decade ago Bousso extended the Bekenstein-Hawking entropy bound into a 
covariant entropy conjecture \cite{Bousso:1999xy}.  While Bousso sharpened the 
definition of the interior of the surface in which the entropy is limited by the 
enclosing area, in his derivation he also made the crucial assumption that there 
should be {\em no singularities} in the interior in question.  (Cf. page 9 of 
\cite{Bousso:1999xy}.)  Bousso also pointed out that "Because the conjecture is 
manifestly time reversal invariant, its origin cannot be thermodynamic, but must 
be statistical. It thus places a fundamental limit on the number of degrees of 
freedom in nature."

In case of black holes, for example, the non-singularity assumption appears to be an obvious necessity.  Otherwise one can imagine white holes channeling entropy into the volume that is independent of the surface area of the black hole.  This could easily violate the entropy bound.

\subsection{Time reversal thinking on the number of states problem}
It is natural to get an idea of our problem for the ``entropy principle''
by thinking for a moment in the time reversal way:

If we think about how one of the to a volume-behaving entropy enourmously
many states could have come about, we may produce the answer by thinking 
time-reversed: What would happen if we started with a typical 
state taken out of the situation with a volume proportional entropy,
and reversed the Hubble expansion to be a Hubble contraction.
That would mean a situation with a very high energy density over a very 
large extension and would of course correspond to a world that were already 
to be considered inside a black hole. Also it would have already so much 
entropy that it would be too much for a/one  black hole. Rather what 
such a system would develop into would be many many black holes. As such a 
collapsing universe with a lot of energy density develops the energy 
density gets even bigger and after some time there will be many relativly 
small subregions which have both too much energy and too much entropy
to avoid being black holes. So at some stage it would develop into an 
approximately smooth distribution of ``small'' black holes. This would mean
a kind of piecewise collaps - even before the naively calculated total 
collaps, when the general size of this universe would go to zero. One could 
say that this in naive sense calculated collaps due to the radius going 
to zero  never gets realized, because the piecewise collaps into 
seperate black holes takes over effectively and forms a collaps at an 
earlier stage. 

Now time reversing this scenario back to the real world, its means 
that the majority of the to the volume behaving entropy corresponding 
states are of such a nature, that they could only be formed from an
earlier stage of the Universe containing enourmously many ``small''
``white holes'' rather than comming from a genuine Big Bang or other 
single or few singularity picture as usual cosmology tells.

Since the ``white holes'' - meaning as we just used here the time 
reversed black holes - are precisely the most important example of what 
a cosmic cencorship principle should forbid, it is clear that the 
majority of states in the volume-based entropy scenario are  
{\em cosmic censorship forbidden} states. In this way there is at least 
the hope that it is the cosmic censorship that can bring the number of 
states down to match the Bekenstein-Hawking-area law.

\section{Can we Rescue the Cosmological Constant 
Derivation?}\label{resolution}
At first there might seem to be a chanse to rescue the work 
of one of us on deriviving the cosmological constant being small
derivation by just saying:

Taken as simply an ultraviolet cut off straight away it looks dangerous 
for locality with the correlations in the cut off needed to make the cut 
off match the entropy principle. However, if we now interprete the cut 
off as mainly due to the initial conditions occuring due to say cosmic 
censorship requierements it would sound much more acceptable, since it 
would no longer even threaden the locality in a genuine sense. 
You cannot really in a world, which has come from a development, in which 
a cosmic censorship principle were valid, send messages faster than light
or the like. At least we have no experimental evidence against that 
we should live in a world with no white holes (unless though perhaps 
Big Bang itself 
should be considered a white hole or a cosmic censorship violating event).
So a cut off considered due to a cosmic censorship would seemingly not
be against what we could believe. However, then the problem would be: 
The main job of the cut off which we should obtain due to the entrpy 
principle were to limit the zero-point energy of the quantized field 
theory of the world, say of the Standard Model. At first one would think
that the cosmic censorship and other agents that could influense the initial
state would not really influence the zero-point energy! One would say this 
because when we think of initial conditions caused by such influences as 
cosmic censorship or from inflations and development whatever, then one has 
in mind that all those high frequency modes which are at a certain time 
not excited by onshell particles will fluctuate nevertheless ``peasefully''
in their zero-point fluctuation way. We so to speak normally imagine 
that the zero point fluctuations for the high frequency modes just are there
as in vacuum for all the frequencies higher than the ones relevant for the 
state being realized. In this philosophy the initial state and thus 
the cosmic censorship would not get true access to influence the 
cutting off of the high frequency modes. If we cannot get the 
cosmic censorship influence the higher frequencies zero mode fluctuation
of course the above discussion and proposal to use cosmic censorship
would not help. 

\subsection{But could initial state effects possibly influence zero-point 
fluctuations?}
But now really the question is:
Shall we take it for a good argument that zero-point oscillations of 
very high frequencies are organized to be present as soon as we reach 
temperatures where they are no longer excited? At first one would again 
say: yes, it is reasonable that the high frequency modes would fall to 
their zero point fluctuation level but no longer as the Universe expands 
with a very strong Hubble expansion and effectively the excitation 
of a mode is moved from one mode to a lower one due to this expansion.
The zero point oscillation cannot be reduced by the Hubble expansion 
and the higher frequency modes would seemingly have to stay in their 
zero point fluctuation. 

But are we not more and more dreaming about a phantacy world of 
high frequencies which never according to the entropy principle 
should even have a chanse to be realized? If one turned the 
philosophy a bit around one would say: We have this phantastic dream of 
there existing a number of possible states of the Universe system 
which is the number of states corresponding to an entropy going 
with the {\em volume} of the Universe, but on the other we know from entropy
principle or essentially equivalently from the cosmic censorship that 
it is only a very tiny minority of these states that have a true chanse to 
be realized. In a way it would be most sensible if in an ontological 
way only the states that have at least the chanse corresponding to the 
entropy principle would exist in a sense of being present in the 
most fundamental theory. But if that were so then all these phantasy states 
making up the to volume proportional entropy ought not to be there.
That would of course make it more strange to worry about the zero point 
energy involved in the many high energy modes which can essentially
ever be excited, or at least almost never. It should namely 
be had in mind that at colliders like LHC we do excite very high modes 
which are normally - i.e. in the almost empty universe - never excited.

Shall we really imagine that in the fundamental theory at the ontological
level which we shall may be once find the degrees of freedom relevant 
for the LHC are in some strange way being built up from some degrees 
of freedom that at first looked like being made for a bit microwaves 
in a low frequency passing far behind the Moon? Shall we really imagine 
that ontologically at the end the degrees of freedom are being shuffled
around so that when the LHC needs some more degrees of freedom it collects 
them up from perhpas big distances away ? Although it sounds a great 
challenge to construct just a model showing that such an idea is possible 
in a local way, it may not be totally excluded since either a clever way 
may be found or nature might at the root of it not respect our usually 
expected principles of locality.
\section{Conclusion}
\label{conclusion}
We have discussed some problems with the model of one of us 
solving the cosmological constant problem
- of the surprisingly small size of the cosmlogical constant 
found experimentally - by using the entropy principle 
(of the entropy only going as the surrounding area). 
The major problem is really a problem with the entropy 
principle rather than only with the proposed solution to the 
cosmological constant. You namely cannot interprete the 
entropy principle at all as a restriction given on the number 
of states as due to some conventional cut off. So either you 
must say that the entropy principle has nothing to do with 
the number of states allowed by an ultraviolet cut off - but
is say a question of the initial state only 
(perhpas via cosmic censorship)- or we must be satisfied by 
an ultraviolet cut off that at least at first looks rather 
complicated with one would say mysterious correlations.
It may be that these ``mysterious correlations'' could sound 
sensible from a speculative fundametal physics point of view.

\subsection{Outlook and hope}
It looks that our discussion is driving us in the direction 
of asking how much reality there is at the fundamental level
in the zero point fluctuations of the various fields. For instance
in last years discussions there were a contribution by one 
of us(H.B.N.) Moultaka and Nagao and Norma Mankoc Borstnic\cite{B10}
related to the quantum mechanics philosophy going back to De Broglie.
The crux of the matter is that the quantum system {\em has} a
position even when it is not in a position eigenstate! Translated into 
field theory we might take this to maean that the fields {\em have}
values even when they are not in an eigenstate field values. This is of 
course crazy and in disagreement with Heisenberg uncertainty principle,
but for Bohm and De Broglie the philosophy is different. If we bought 
the theory of Bohm and De Broglie for fields we would not have to 
believe that there were truly(ontologically) zero point fluctuations,
but could leave it as a more difficult physical question to be 
answered by a deeper understanding of the more fundamental theory 
we are looking for. But to by such a means get the contribution 
to the cosmological constant from the high frequency modes be negligible
as is hoped to get rid of the cosmological constant problem it would 
e needed that this hoped for theory behind (or at the end at the most 
fundamental level) would put the energy of the high frequency modes to 
be lower than what is possible in quantum mechanics with its minimum at the 
zero point enrgy for a harmonic oscillatior. We would have to put the 
harmonic oscillator corresponding to the high frequency modes of the 
fields - for which we want to get rid of the contribution to the 
cosmological constant - to have {\em both} zero momentum and zero 
position rather exactly! For Heisenberg impossible, but for Bohm and 
De Broglie the question is more to be studied with more details added.

\subsection{The more Private hope using the Complex Action Model}   
Since as the discussion above has shown the proposed model for solving 
the cosmological constant problem has been threadened - although not
definitively killed on that ground - by locality principle. If it should 
at the end turn out to be indeed needed to give up such principles and 
for instance go to a model like the ``complex action model'' by one of us 
(H.B.N.) and Ninomiya - originally based on ideas developped 
by H.B.N. and Don Bennett - which were the model used in the above 
mentioned discussion contribution from last year by Moultaka et. al.
we might use such a model to suggest what should be the classical 
values of the high frequency fields. In other words we might now
aks in the complex action model for how the in this model essentially 
classically standing fields behave (it means they do not respect 
Heisenberg in the way we ask for their fundamental values). 
We can almost immediately guess the answer: In the complex action model 
the guiding principle is that the initial conditions get set so as to 
minmize the imaginary part of the action. Now this must looking at 
world as it is mean that to have the vacuum we live in is extremely 
favourable to lower this imaginary part of the action. Then presumably 
if the ``God''(having such a minmization principle arranging 
the quantity the imaginary part of the action $S_I$ to be minimal
is almost like having a ``God'' in quotation marks governing 
the world to make ``His'' deficit $S_I$ as small as possible,
preferably negative)   behind the governing of the Universe were so
keen to make so much vacuum, ``He'' should be even more keen to 
push the vacuum the last little bit by putting the fields that in our usual 
vacuum pictuture are in their zero point fluctuation states the last bit 
so as to have both momentum and position go to the bottom. If the imaginary 
part of the action $S_I$ is just a reasonably smooth function(al) of the 
field configurations and their conjugate momenta and it seems that the 
most beloved state ( by ``God'', meaning giving the most favoured 
meaning low $S_I$) is the vacuum then if it were possible almost 
certainly the classical replacement for the vacuum having the fields
exactly zero would have an even lower $S_I$ and thus be even more 
beloved! So our complex action model would indeed predict that the 
values of the fields - only being allowed by De Broglie and Bohm -
would be so that the zero point enrgy would be killed at the fundamental 
level! That would as the reader can immediately understand be wonderful
for the cosmological constant model we have discussed: we suggested in the 
last years discussion that the complex action model could function 
approximately as a model behind the Bohm-De Broglie picture, and now that 
the prediction from complex action would then be that the vaccum fields 
would be - at least when not too much disturbed - be put to zero exactly
(contrary to what Heisenberg uncertainty would allow, but tht is o.k. 
in Bohm De Broglie and in complex action interpreted the right way 
as being ``by hindsight'', i.e. including knowledge collected by a 
measurement) as well as the conjugate momentum to the field modes in 
question. 
      
Now using the complex action might however be an almost too high price 
in the sense that Ninomiya and one of us (H.B.N.) already have an
article suggesting that this complex action model is good for helping
with the cosmological constant problem \cite{ccNN}. In the kind of thinking 
in the articles seeking to solve cosmological constant problem in the 
complex action model or related models previously the philosophy were 
however quite a bit different in as far as in that sort of works it would 
rather be assumed that there is presumably very big {\em bare cosmological
constant}, which simply gets adjusted by essentially the already mentioned
``God'' in quotation marks so as to minmize the imaginary part of the 
action. If ``He'' for some reason should want a Universe avoiding collabs 
but not expanding faster than necessary ``He'' could easily arrive to 
vote for a small cosmological constant. But if ``He'' has power to 
adjust the {\em bare } cosmological constant, ``He'' hardly need to
for that reason go into adjusting zero-pointfluctuating modes, but 
``He'' according to the above does it anyway.          
\section*{Acknowledgement}
One of us (H.B.N.) wants to thank Keiichi Nagao Ibaraki University 
for discussions about the problems of number of states relativelong 
time ago at a conference in Holbaek.

\end{document}